\documentclass{icrc2009}

\usepackage{graphicx}   
\usepackage[caption=false]{caption}    
\usepackage[font=footnotesize]{subfig} 
\usepackage{fixltx2e}
\usepackage{url}

\newcommand{\shorttitle}[1]%
{\markboth{Proceedings of the 31\MakeLowercase{$^{st}$} ICRC, {\L}\'{o}d\'{z} 2009}{#1} }
\newcommand{\etal}{\MakeLowercase{\textit{et al. }}} 

\begin{document}
\title{A Flexible High Demand Storage System for MAGIC-I and MAGIC-II
using GFS}

\author{\IEEEauthorblockN{E. Carmona\IEEEauthorrefmark{1},
			  J. A. Coarasa\IEEEauthorrefmark{2} and
                          M. Barcel\'o\IEEEauthorrefmark{3} for the
                          MAGIC Collaboration}
                            \\
\IEEEauthorblockA{\IEEEauthorrefmark{1}Max-Planck-Institut f\"ur
  Physik, Munich, Germany}
\IEEEauthorblockA{\IEEEauthorrefmark{2}University of
  California, San Diego, La Jolla, California, USA}
\IEEEauthorblockA{\IEEEauthorrefmark{3}Institut de
F\'{\i}sica d'Altes Energies, Barcelona, Spain}}

\shorttitle{E. Carmona \etal MAGIC Data Storage with GFS}
\maketitle

\begin{abstract}
MAGIC is currently the Imaging Cherenkov Telescope with the
worldwide largest reflector currently in operation. The initially
achieved low trigger threshold of 60 GeV has been further reduced by
means of a novel trigger that allows the telescope to record gamma ray
showers down to 25 GeV. The high trigger rate combined with the 2 GHz
signal sampling rate results in large data volumes that can reach 1
TByte per night for MAGIC-I and even more with the second MAGIC
telescope coming soon into operation. To deal with the large storage
requirements of MAGIC-I and MAGIC-II, we have installed the
distributed file system GFS and a cluster of computers with concurrent
access to the same shared storage units. The system can not only
handle a sustained DAQ write rate above 1.2 kHz for MAGIC-I, but also
allows other nodes to perform simultaneous concurrent access to the
data on the shared storage units.  Various simultaneous tasks can be
used at any time, in parallel with data taking, including data
compression, taping, on-line analysis, calibration and analysis of the
data. The system is designed to quickly recover after the failure of
one node in the cluster and to be easily extended as more nodes or
storage units are required in the future.
\end{abstract}

 
\section{The MAGIC Experiment Data Requirements}

The MAGIC telescope belongs to the latest generation of imaging
atmospheric Cherenkov telescopes (IACTs) characterized by a high
sensitivity. The MAGIC telescope is also the IACT with the lowest
energy threshold, 60~GeV, thanks to the large mirror diameter (17~m)
and the high quantum efficiency photomultipliers. The low energy
threshold of the telescope can be further reduced when the so-called
{\em Sum Trigger}~\cite{MAGIC2} is used. Thanks to this trigger,
the MAGIC telescope has been able to reach an energy threshold of only
25~GeV and has detected for the first time the VHE pulsed emission
from the Crab Nebula pulsar~\cite{CrabPulsar} above 30 GeV.

The MAGIC low energy threshold results in high event rates that impose
demanding requirements on both the data acquisition system (DAQ) and
the data storage devices. The DAQ of the MAGIC
telescope stores the complete waveform from all the 577 pixels in the
camera. The sampling is done using 10~bits and a sampling rate of
2~GHz~\cite{DAQ}. A total of 50 samples per event and pixel are
stored. The total size of an event is 60.7~kByte and the average
trigger rate is around 350~Hz. The average raw data
volume is around 21~MByte/s. Sum trigger observations can increase the
trigger rate above $\sim$1~kHz. In that case, the raw data volume
reaches 60~MByte/s and one night of observation can easily exceed
1~TByte of data.

The DAQ is only one of the processes with high demanding requirements
in terms of storage. Other processes executed at the telescope site
need access to the written data and also have important I/O
requirements. During observation, the on-line analysis program
produces a fast analysis using the data just being taken. To do so,
the on-line analysis software accesses the raw data a few seconds
after they have been written to the disks.  Events are processed at a
speed similar to that of the DAQ, thus requiring a reading performance
comparable to the writing performance needed by the DAQ.

The high data volumes generated by the MAGIC telescopes cannot be
transferred directly through network to the
datacenter~\cite{Datacenter}, but have to be processed first. Data are
first compressed to reduce their volume in disk (factor $\sim$3.1
reduction) and later taped. Several CPU cores are used in parallel to
perform the compression of the data, so the speed of the compression
process is usually limited by the I/O speed of the disks where the raw
data are stored. Taping is done to have a backup copy of the data and
to transfer the data to the datacenter. The taping process requires a
speed up to 60~MByte/s (LTO3 tapes used for MAGIC-I).

In order to make data available as soon as possible, the on site
analysis software performs the calibration of the data
also at the telescope site. Calibrated data (factor $\sim30$ reduced
in comparison to raw data) can hence be
 easily transferred to the datacenter using a fast network
connection. Typically, calibrated data is available at the datacenter
less than 24 hours after the data taking is finished. Calibration and
data transfer are two of the most demanding processes in terms of I/O
performance. The read and write disk speed must be as high as possible
in order not to be limited by the disk access. These processes are
also quite demanding in terms of CPU and typically several instances
of the on-site analysis and the data transfers must be executed in several
CPU cores.

The second MAGIC telescope~\cite{MAGIC2} (currently under
commissioning) is equipped with 1039 pixels and the DAQ works also
with a sampling rate of 2~Gsamples/s. As a result, the event size
increases to $\sim$110~kByte, doubling the expected data
volumes. Similar processes as those described for the first MAGIC
telescope have to be executed for the data coming from the second
MAGIC telescope, the difference being the data flow and the CPU needed
to process the double amount of data. Although the storage systems for
MAGIC-I and MAGIC-II data can be completely independent, a system
where both telescopes are integrated would allow us to dynamically
choose which resources (CPUs or storage) are to be used in each
telescope.

Many of the MAGIC requirements can be provided by a Storage Area
Network (SAN) solution. A SAN is an architecture to attach remote
computer storage devices to servers, such that the devices appear as
locally attached to the operating system. This can provide the
required high I/O performance but does not solve the problem of
offering a consistent view of the file system to all computers
directly attached to the storage devices. In addition, the required
solution should also be flexible and scalable in order to accept the
changes that will be introduced when the second telescope starts its
operation.

\section{The Red Hat Cluster Suite and the Global File System Filesystem}

When two or more computers work together to perform the same task they
form a cluster. Clusters of computers are built to provide storage,
high availability, load balance or high performance. In the case of
MAGIC, a cluster of computers is used to provide a consistent file
system image across the servers in the cluster. All members of the
clusters (nodes) have simultaneous read and write access to a shared
file system. The storage cluster solution used in MAGIC uses the Red
Hat Cluster Suite (RHCS)~\cite{RHCS} and the Red Hat Global File System
(GFS)~\cite{GFS}. 

The RHCS is an integrated set of software that can be configured to
suit the needs of each cluster system. RHCS provides the basic
functionality for the nodes to work together: configuration-file
management, membership management, lock management and fencing. This
functionality is provided by the different daemon process executing in
all nodes: CCSD, CMAN, LOCKD and FENCED.

\begin{itemize}

\item{CCSD} manages the cluster configuration, making sure that all
  cluster members have the same cluster configuration. Changes in the
  cluster configuration can be propagated to all nodes by the CCSD. 

\item{CMAN} manages cluster quorum an cluster membership. CMAN keeps
  track of cluster quorum by monitoring the number of cluster
  nodes. When more than half of the nodes are active the cluster
  reaches quorum. When the cluster does not have quorum all cluster
  activity is blocked, preventing the cluster fragmentation that could
  lead in a storage cluster to data corruption in the file
  system. Membership is tracked by monitoring the messages that the
  nodes exchange. These messages are communicated between nodes via
  Ethernet.

\item{LOCK} is a service that provides the mechanism to synchronize
  access to shared resources. In the RHCS the Distributed Lock Manager
  (DLM) is the lock manager. It runs in each cluster node, distributing
  the lock management across all nodes in the cluster. 
 
\item{FENCED} manages the disconnection of a node from a cluster
  shared storage, cutting I/O from shared storage and ensuring data
  integrity. When a CMAN process determines that a node has failed
  this information is communicated to all nodes. When notified, the
  fenced daemon fences the failed node. Different fencing methods
  exist to guarantee that the fenced node does not have access to the
  shared storage. Two of the most widely used fencing methods are {\em
    Power fencing} (power off the failed device), {\em Fibre Channel
    switch fencing} (close the Fibre Channel switch port of the
  failed node).

\end{itemize}

\begin{figure}[!t]
  \centering
  \includegraphics[width=3.2in]{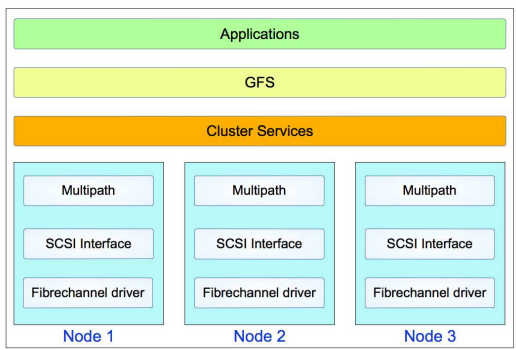}
  \caption{Software layers involved in the SAN storage solution with
    GFS implemented for MAGIC.}
  \label{fig: software-stack}
\end{figure}

On top of the cluster services the Red Hat Global File System (GFS)
can be installed (see figure~\ref{fig: software-stack}). GFS is a
native file system that interfaces directly with the Linux kernel file
system interface. Although GFS can work in a stand alone node, it
shows all its capabilities when implemented as a cluster file
system. GFS provides data sharing among GFS nodes in a cluster,
providing a single, consistent view of the file system. With GFS,
applications running on different nodes can share files in the shared
storage in the same way as applications running in the same machine can
share files in a local file system (figure~\ref{fig: software-stack}).

Finally, GFS can be deployed in a variety of configurations that can
fit any user needs. This is an open source solution supported by
different widely used Linux distributions like RHEL or Scientific
Linux. In addition, this solution is also highly flexible and can be
easily scaled as new needs appear.

\section{Implementation in MAGIC}

MAGIC has implemented at the telescope site a SAN solution with Fibre
Channel technology that uses RHCS cluster suite and the GFS file
system. At present seven nodes running Red Hat Enterprise Linux (RHEL) 4.3
and Scientific Linux (SL) 4.4 are members of the cluster. At a low
level, the connection between the cluster nodes and the storage
devices is done using Fibre Channel technology.

At the moment MAGIC has installed 4 RAID (Redundant Array of
Independent Disks) systems with capacities between 7 and
21~TBytes. The RAID units consist of several physical SATA disks
configured in RAID5 (redundancy based on parity data distributed
across all member disks), which offers redundancy without sacrificing
much of the I/O performance. Maximum read/write (R/W) speed of the
RAID systems is between 250 and 300 MByte/s. The RAID systems are
equipped with two Fibre Channel interfaces each.

\begin{figure*}[th]
  \centering
  \includegraphics[width=6in]{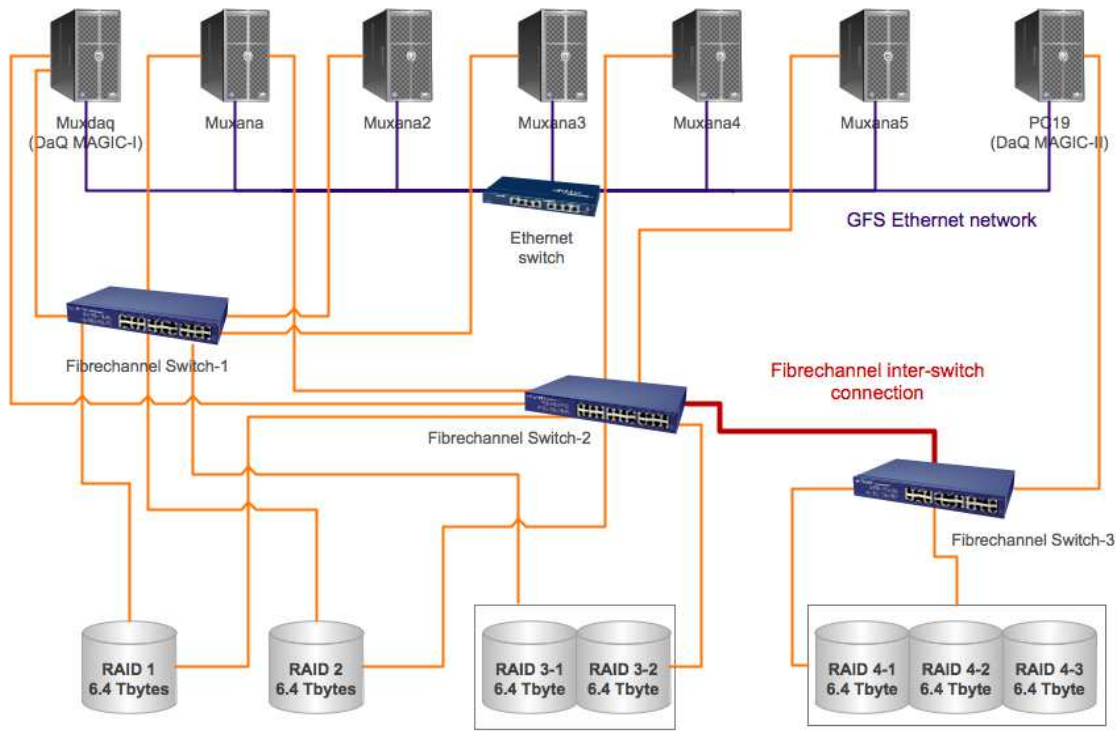}
  \caption{Topology of the Fibre Channel storage network. Most nodes
    are equipped with one single Fibre Channel interface but all RAID
    systems are equipped with two interfaces to achieve higher
    reliability and performance. Cluster messages among the nodes are
    exchanged through a dedicated ethernet network.}
  \label{fig: GFS-topology}
 \end{figure*}

The speed of the Fibre Channel connection is 2 or 4~Gbps (250 and 500
MByte/s) for the different RAID systems and nodes. The connection
between nodes and RAID systems is not direct. All devices in the
storage network are connected to Fibre Channel switches that are
configured to work in a fabric topology. A total of three switches are
used in our current configuration to guarantee the access to the
storage devices in case of a switch failure. The switches can be
cascaded if a special license is purchased. Currently only two of the
switches are connected (see figure~\ref{fig: GFS-topology}).

The RHCS-GFS suite has been installed in all nodes as well as the
tools and drivers needed for the Fibre Channel connections. Currently
the system is running using the same RHCS and GFS versions installed
in two different Linux distributions and different kernel versions.
The software structure is organized as follows. At the lowest level
the Fibre Channel driver connects the nodes to the switches and the
storage devices. On top of that, the operating system can access the
devices as SCSI devices directly attached to the node. The multipath
service combines all different routes to the same device under a
common device. On top of this, the cluster services of the RHCS run on
each node providing all services needed for the GFS layer. The GFS
provides the consistent view of the file system to all nodes in the
cluster, mounting the device provided by multipath and
formatted with the GFS file system. On top of the GFS the different
applications are executed in each node.  

The storage cluster is currently composed of seven nodes. Due to GFS layer
the maximum R/W speed in the RAID systems is limited to only
$\sim$200~MByte/s for a single RAID system. Different nodes can have
access at the highest rates to different RAID systems
simultaneously. The RAID systems with higher capacities are subdivided
in partitions with the same size as the smallest RAID systems
($\sim6.4$~Tbytes/partition). This size was limited by the first GFS
version installed in the system. The seven currently installed nodes 
contain 42 cores that have direct access to storage devices and share
the same consistent view of the file system. All messages exchanged by
the CMAN daemon are transmitted through a dedicated 1~Gbps ethernet
network and a dedicated ethernet adaptor in each node.

\section{Reliability, Flexibility and Scalability}

Additional aspects that were considered when adapting this solution 
include the reliability, the flexibility and the scalability of
the system. The GFS solution implemented in MAGIC offered excellent
characteristics in all these aspects.

\subsection{Reliability}

At the lowest level, the RAID units are configured in RAID5 which
guarantees that no data is lost when one of the disks in the RAID
units fails. In addition, the control software of the RAIDs 
automatically uses the units configured as spares in the RAID when a
disk failure occurs.

At the Fibre Channel level, a failure in one of the connections or in
a full switch can be handled by the multipath daemon. When a node or a
storage device has more than one Fibre Channel interface, different
routes exist to access the same device. The multipath daemon unifies
all possible paths to a storage device as one unique device. In case
of a failure, multipath automatically routes the traffic using the
working paths without any action required. At present three Fibre
Channel switches are used also to provide higher reliability. 

Node failures are handled by the CMAN service, part of the RHCS. When
a node failure happens the FENCE service blocks all nodes access to
the shared storage units until the non-responding node is
blocked. This is done to prevent data corruption in the file
system. The fencing of a node is done in our implementation by closing
the ports in the Fibre Channel switches where that node is
connected. This guarantees that the non-responding node cannot access
the RAID disks. After the node is fenced, the access to the RAIDs
system is enabled again for all members of the cluster. The time
needed for this operation is of only a few seconds and it is automatically
handled by the CMAN and fence services.

\subsection{Flexibility}

Cluster configuration is contained in a single file that is identical
in all nodes belonging to the cluster. Any change in the cluster
configuration (like adding a new node) can be easily propagated to all
nodes in the cluster with the CCSD daemon. 

Since all nodes have direct R/W access to the GFS file systems,
applications running in any node can choose where to read or write
data. This allows for the dynamical choice regarding how many applications are
needed for each task and where they have to be executed, and provides a
high flexibility that could not be reached with other solutions.

\subsection{Scalability}

This system can be easily extended by adding either more nodes or more
storage devices. At present, less than 30\% of the Fibre Channel
switch ports are used. The Fibre Channel devices use the
fabric topology making it possible to add more nodes or storage
devices just by plugging them to a new port in the switch.

Adding more nodes to the storage cluster is also not a problem. At
present seven nodes are connected and we have plans for adding at least five
more new nodes to the cluster. The only limitation when adding more
nodes is that the GFS file system must have a number of journals equal
or larger than the number of nodes connected. That number is
configured when the GFS partition is formatted. GFS supports over 100
nodes.

\section{Conclusions}

The high demands that the data taking and the data process impose on
the on site storage systems used in the MAGIC telescopes have been
addressed using the RHCS and the distributed filesystem GFS. This
Linux open source solution offers a consistent view of the file system
and high I/O performance since all nodes in the cluster have direct
access to the storage devices. It also provides high reliability and
high flexibility that allows us to allocate resources dynamically
where they are more needed or to change the configuration if a failure
in a node or a storage device occurs. The system is also designed to
provide easy scalability in the midterm future when new needs appear
in the MAGIC-II phase of the experiment.


\end{document}